# Generation of Ultra-intense Gamma-ray Train by QED Harmonics

Chen Liu,[1] Baifei Shen,[1,2,*] Xiaomei Zhang,[1,*] Liangliang Ji,[1] Wenpeng Wang,[1] Jiancai Xu,[1] Xueyan Zhao,[1] Longqing Yi,[1] Yin Shi,[1] Lingang Zhang,[1] Tongjun Xu,[1] Zhikun Pei[1] and Zhizhan Xu[1]

[1]*State Key Laboratory of High Field Laser Physics, Shanghai Institute of Optics and Fine Mechanics, Chinese Academy of Sciences, P. O. Box 800-211, Shanghai 201800, China*

[2]*IFSA Collaborative Innovation Center, Shanghai Jiao Tong University, Shanghai 200240, China*

Abstract:

When laser intensity exceeds $10^{22}$ W/cm$^2$, photons with energy above MeV can be generated from high-order harmonics process in the laser-plasma interaction. We find that under such laser intensity, QED effect plays a dominating role in the radiation pattern. Contrast to the gas and relativistic HHG processes, both the occurrence and energy of gamma-ray emission produced by QED harmonics are random and QED harmonics are usually not coherent, while the property of high intensity and ultra-short duration is conserved. Our simulation shows that the period of gamma-ray train is half of the laser period and the peak intensity is $1.4\times10^{22}$ W/cm$^2$. This new harmonic production with QED effects are crucial to light-matter interaction in strong field and can be verified in experiments by 10PW laser facilities in the near future.

The source of ultra-short, extremely high frequency radiation has been successively pursued for its significant applications in ultrafast physics and attosecond science [1, 2]. Motivated by this prospective application, high-order harmonic generation (HHG) from laser-plasma interaction has become a dynamic field. The physical mechanisms of high-order harmonic generation are related to the laser intensity or the normalized laser amplitude $a = eE/m_e c\omega$. Here $E$ is electric field of the laser, $\omega$ is laser frequency, $c$ is the light speed, $e$ and $m_e$ are electron's charge and mass, respectively. The radiations from HHG under different laser intensities show different characters. When it reaches $10^{11}$ W/cm$^2$, the external field imposed by laser is still much weaker than the atomic Coulomb field. Frequency up-conversion of laser pulse including frequency doubling and tripling via nonlinear crystals can produce ultraviolet (UV) radiation and these processes may be the early studies of high-order harmonic generation. Nonlinear interactions taking place under this parameter range is referred as regime of perturbative nonlinear optics. When the laser intensity has the order of $10^{15}$ W/cm$^2$ ($a \ll 1$), due to tunnel ionizing and radiation by recombination of electrons, coherent XUV to soft X-ray radiation are produced in the interaction of such laser with noble gases [3-5]. The cutoff energy of radiation can reach several hundred eV. Furthermore, HHG enters the relativistic regime as laser intensity increases to $10^{18}$ W/cm$^2$ ($a \sim 1$). The impinging of relativistic laser pulse on solid targets produces X-rays with cutoff energy exceeding keV and the oscillation of the target surface contributes to the harmonic generation. This model is known as relativistically oscillating mirror (ROM) [6, 7], which has been confirmed in experiments [8-10]. However, the HHG process will be drastically changed if the energy of emitted photon exceeds $m_e c^2$ (the rest energy of an electron) and thus the radiation reaction force cannot be neglected [11, 12]. In this case, QED effects [13-18] become important and strongly influence the radiation process [19]. According to QED theory, the typical energy of emitted photons can be estimated by

$hv = 0.44\eta\gamma m_e c^2$ [20], here $\eta = \gamma|E_\perp + \beta\times B|/E_{cr}$ is QED parameter, $\gamma$ and $\beta$ are the Lorentz factor and velocity of electrons normalized by the light speed, $E_\perp$ is laser electric field perpendicular to the velocity of electrons and $B$ is laser magnetic field, $E_{cr} = 1.3\times 10^{18}$ V/m is the Schwinger field. Because of $\gamma \sim a$, the QED effect begins to become important when $a > 100$. For a laser with wavelength of 800 nm, the corresponding intensity is above $10^{22}$ W/cm$^2$. In the QED regime, two new features appear in the high-order harmonic generation. For one thing, the radiation process of charged particles is no longer continuous but one photon after another due to quantum effects. And the occurrence of radiation is completely random and determined by probability in contrast with the classical one. For another, since the phases of emitted photons are not locked and their wavelengths are extremely short, the radiation is hardly to be coherent but appears the characteristic of a particle beam. In addition, the changing of the trajectory due to radiation reaction force also affects the relative phases of emitted photons. In this sense, we term HHG in this regime as "QED harmonics" in order to distinguish it from the classical coherent HHG process.

In this letter, we firstly report the generation of an attosecond gamma-ray train based on QED harmonics from the interaction of an ultra-intense laser with a solid target.

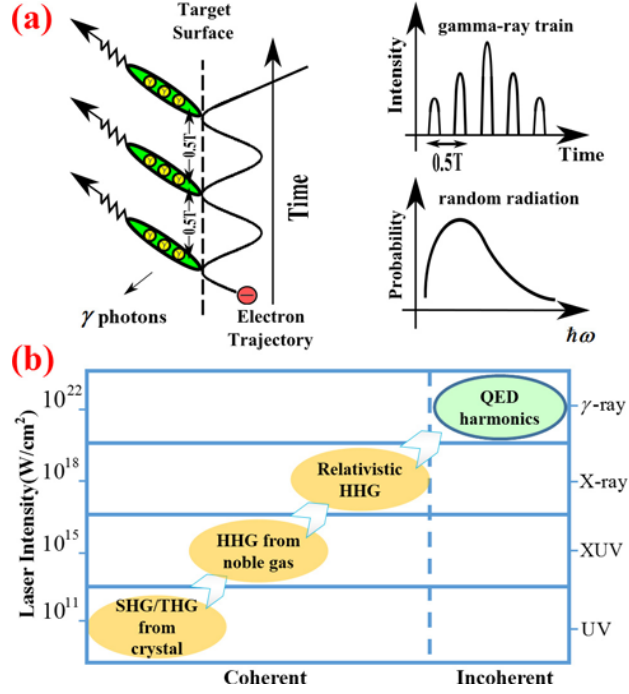

FIG. 1 (color online). (a) The scheme of gamma-ray train generation based on QED harmonics. (b) The evolution of HHG in different parameter regimes, including SHG and THG from crystal in perturbative nonlinear optics regime, HHG from noble gas in strong field nonlinear optics regime, relativistic HHG in the regime of relativistic optics and QED harmonics from QED optics regime.

The typical scheme of gamma-ray train generation from QED harmonics is shown in Fig. 1(a). The radiation process is studied with one and two-dimensional particle-in-cell simulations performed by the code EPOCH [21, 22]. A linearly p-polarized pulse, with wavelength $\lambda = 0.8$ μm and peak amplitude $a_0 = 500$ ($I = 5.35 \times 10^{23}$ W/cm$^2$), propagates along the $x$ direction illuminating normally on a gold foil. The laser pulse has a profile of $a = a_0 \sin^2(\pi t/2\tau_0)$, where $\tau_0 = 5T$ is the pulse duration (FWHM), $T$ is the laser period. The density of target is $n_e = 500 n_c$, where $n_c = m_e \omega^2 / 4\pi e^2$ represents the critical density. The simulation box is $30\lambda$ in the $x$ direction, and the target is initially located in the region of $10\lambda < x < 20\lambda$ and represented by 1000 macro-particles for each species per cell. Each cell occupies a size of $0.01\lambda$. The ions are considered to be fixed first to simplify the situation and to get a clear understanding of the underlying physical mechanism.

The density distributions of both electrons and photons are analyzed in the spatial-temporal plane (*x-t*), as shown in Fig. 2(a) and Fig. 2(b). One can see that the laser pulse cannot penetrate into the overdense target and at the front surface an electron layer is formed, which oscillates periodically due to the imbalance of laser ponderomotive force and electrostatic force. This is similar to the case of conventional relativistic HHG. Comparing the evolutions of the photon and electron density in Fig. 2, one can see that the gamma-ray emission comes from the front surface of the target, where the relativistic electrons serve as the radiation sources and emit energetic photons as they oscillate violently in the layer. Fig. 2(b) also shows a clear periodic structure of the gamma-ray photons. This can also be seen in Fig. 3(a), where the gamma-ray train observed at the position $x=5\lambda$ is shown. Here, only photons with energy above 1 MeV and within a small angle around *x* axis are included. ($px/py>7$, where $px$, $py$ are the momentum of photon in the *x* direction and *y* direction, respectively. This corresponds to an angle of $16.26°$.) Fig. 3(a) shows that the period of the gamma-ray train is $0.5T$ and the peak intensity of the gamma-ray train is about $1.4\times10^{22}$ W/cm$^2$.

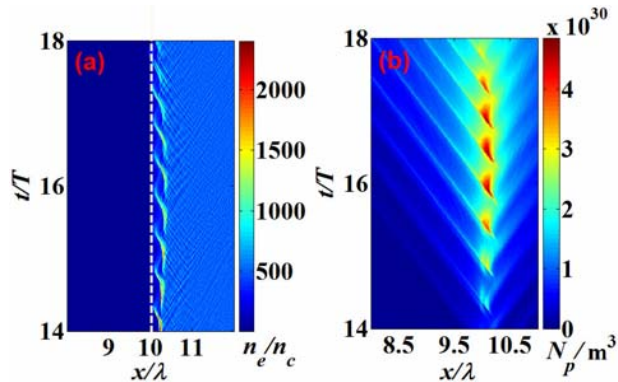

FIG. 2 (color online). The density distributions of electrons (a) and photons (b) in the spatial-temporal plane (*x-t*). The lateral axis shows different positions of the plasma target and $x=10\lambda$ (white dotted line) is the initial target boundary. The vertical axis gives the evolution time from $14T$ to $18T$. Here, $15T$ corresponds to the time when the peak of the laser pulse interacts with the target.

As discussed above, the gamma-ray emission comes from the overdense layer surface, where relativistic electrons interact with extremely strong electromagnetic field, thus the dynamics of electrons in this layer is crucial to the radiation process. Fortunately, their trajectories can be traced in the PIC simulation, which is plotted with respect to time in Fig. 3(b). Though the movement of a single electron may not be always periodic, the collective movement of large numbers of electrons shows obviously that the electron layer at the plasma boundary oscillates with a period of $0.5T$. This can give rise to the periodic character of radiation. However, under such an intense laser pulse, QED effects bring strong influence on the radiation process. One of these new features is the randomness of radiation. For a single electron, the occurrence of radiation is no longer certain and continuous but random and by quantum. In addition to that, the energy of emitted photon is uncertain and calculated by sampling from the quantum-corrected synchrotron spectrum. Meanwhile, since the energy of emitted photon becomes so large that the radiation reaction force cannot be neglected. It will change the original trajectory of electrons and the synchronization between different electrons will be destroyed due to randomness of radiation. Therefore the photons will not have the same phase, which leads to an incoherent radiation. And, this incoherent, extremely high frequency radiation is more likely to act as a particle beam rather than a wave. Thus, one can predict that its energy spectrum will have a Maxwell distribution similar to a group of particles.

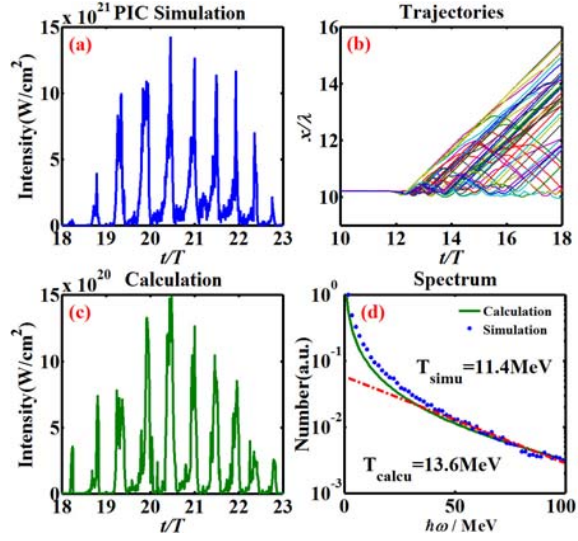

FIG. 3 (color online). (a) The gamma-ray train obtained at $x = 5\lambda$ in the PIC simulation. (b) The trajectories of typical electrons initially located in the same position of the target front. (c) The calculation result of the gamma-ray train by tracing the typical electrons. (d) The energy spectrum of the generated gamma-ray photons. The green line represents the calculation result, the blue dotted line represents the simulation result. The red dashed line is the linear fitting of the photons temperature. The $x$ axis represents the photon energy in the unit of MeV. The $y$ axis represents the number of gamma-ray photons.

To illustrate the particle nature of this gamma-ray radiation, the energy spectrum of emitted photons in our simulation is given in Fig. 3(d). From Fig. 3(d), one can see that the number of photons decreases almost as a linear scale with respect to the photon energy in the semi-log coordinate. This shows that the number of photons may have a distribution of $N \sim \exp(-\hbar\omega/kT)$, which implies that the radiation is incoherent and behaves as a group of particles. This is significantly different from the conventional relativistic HHG process and this indicates a new mechanism is involved in it for which we term as 'QED harmonics'. The linear fitting result (red dashed line) shows that the temperature of emitted photons is about 11.4 MeV. According to QED theory, under the assumption of monochromatic radiation, the average photon energy emitted by electrons is $h\nu = 0.44\eta\gamma m_e c^2$. According to the simulation, $\bar{\eta} \sim 0.18$,

$\bar{\gamma} \sim 300$, the average photon energy equals 11.9 MeV, which agrees well with the temperature from simulation.

In order to verify the simulation result, we treat the trajectories provided in Fig. 3(b) as typical ones of the electrons in the overdense layer and calculate the gamma-ray radiation of these electrons by QED formulae. The total power radiated by a single electron is [23]

$$P = \frac{2}{3}\alpha_f \eta^2 m_e c^2 \frac{m_e c^2}{\hbar} g(\eta), \quad (1)$$

$$g(\eta) = \frac{3\sqrt{3}}{2\pi\eta^2} \int_0^\infty d\chi F(\eta, \chi). \quad (2)$$

Here, $\hbar$ is Plank's constant, $\alpha_f = e^2/\hbar c$ is the fine-structure constant. $F(\eta, \chi)$ is quantum-corrected synchrotron spectrum which can be found in Ref[23]. $\eta$ and $\chi$ are Lorentz-invariant parameters introduced to describe the QED process. They are defined as $\eta = (\gamma/E_{cr})|E_\perp + \beta \times cB|$ and $\chi = (\hbar\omega_\gamma/2m_e c^2)|E_\perp + k \times cB|$, where $k$ is the vector of gamma photon, $\omega_\gamma$ is the frequency of emitted photons. The calculation result in Fig. 3(c) shows a good agreement with the simulation result in Fig. 3(a) except for a smaller value of intensity. This is reasonable because we only calculate the radiation from typical electrons in the target surface. Besides, we calculate the radiation spectrum of the gamma-ray emission from Eqs. (1-2). The spectrum agrees well with the simulation in Fig. 3(d), where the temperature of gamma-ray photons is 13.6 MeV according to our calculation, which is close to 11.4 MeV in simulation.

Considering the effects of ion movement, the situation makes several differences [24, 25]. First, a part of energy transfers to ions from laser pulse and this reduces the energy of electrons and photons since the energy should be conserved. Second, striking by such high intensity laser pulse, ions are pushed along the laser propagating direction (*x* direction). If ions are immobile, the electron layer oscillates around the fixed ions in the target front. Actually, the ions will have a speed along the *x* direction and the electrons move with the ions and oscillate as well. If the speed of ions has the order of the electrons oscillating speed, from Lorentz transformation, one can deduce that the ion movement will enhance the electrons movement in the forward direction (*x* direction) and suppress the movement backwards (-*x* direction) and then the radiation differentiates with the ion-fixed case. Fig. 4(a) shows the gamma-ray emission along the laser propagating direction (*x* direction) and only photons with energy above 1 MeV are included. The peak intensity of gamma-ray train is $8.7 \times 10^{20}$ W/cm$^2$ and the duration of one single pulse is about 264 as. Assuming that the radius of the incident laser spot is 2 μm, the energy of the single gamma pulse is 27.9 mJ and the energy of the laser is 336 J. The conversion efficiency from laser to the single gamma pulse is $8.3 \times 10^{-5}$. The corresponding spectrum of the gamma-ray train is shown in Fig. 4(b) and the temperature of the photons is 5.6 MeV.

Aiming to study whether the multi-dimensional situation affects such process, we perform a 2D PIC simulation in which the parameters of laser and solid target are almost the same with the 1D one. The simulation box is $30\lambda(x) \times 20\lambda(y)$, with $3000(x) \times 2000(y)$ cells and 10 particles per cell. The laser normalized amplitude is $a = a_0 \sin^2(\pi t / 2\tau_0) \exp(-r^2/r_0^2)$, where $a_0 = 500$, $r_0 = 5\lambda$ is the radius of laser spot size, $\tau_0 = 5T$ is the pulse duration (FWHM). The target has a size of $10\lambda(x) \times 16\lambda(y)$, which located from $10\lambda \sim 20\lambda$ in *x* direction and from $-8\lambda \sim 8\lambda$ in *y* direction. The density of the target is $500 n_c$ and the ion is immobile.

In 2D situation, the gamma-ray emission around the *x* axis is shown in Fig. 4(c). One can see that the generation of gamma-ray train still exists. The intensity of gamma-ray train is lower than that in 1D case because the gamma-ray emission has a transverse distribution in 2D case. This can be seen from the density distribution of generated gamma-ray photons at time $t = 17T$ (7 laser periods after the interaction begins) in Fig. 4(d). The observation position is $x = 5\lambda$, so the photons peak propagates to $x = 9\lambda$ at this time ($t = 17T$) corresponds with the pulse signal at $t = 21T$ in Fig. 4(c). In general, multi-dimensional effects have limited influence on the radiation mechanism of QED harmonics and the generation of gamma-ray train.

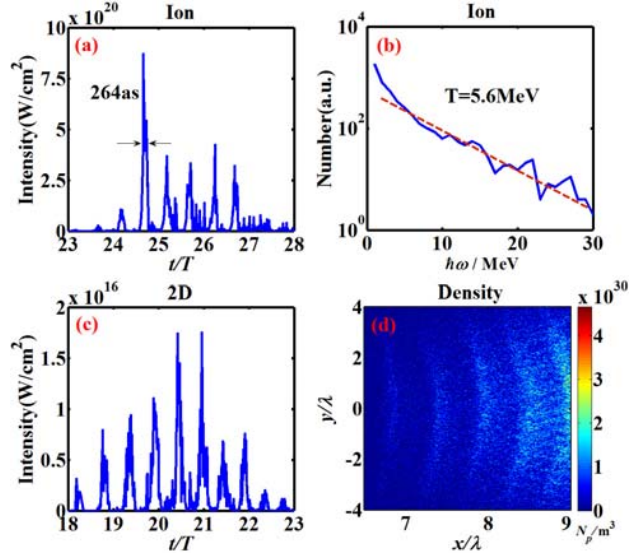

FIG. 4 (color online). (a) The gamma-ray emission in the forward direction obtained at $x = 20\lambda$ when ions are free. (b) The energy spectrum of the gamma-ray train shown in (a) in the case of free ions. (c) The gamma-ray train around the $x$ axis in 2D simulation. (d) A snapshot of the density distribution of generated gamma-ray photons (>1 MeV) at $t = 17T$ in 2D simulation. The photons located at $x = 9\lambda$ corresponds with the pulse signal at $t = 21T$ in (c).

In conclusion, when laser intensity exceeds $10^{22}$ W/cm$^2$, the HHG process from light-matter interaction enters the 'QED regime'. We name high-order harmonics generated in this regime 'QED harmonics'. Different from the conventional HHG, QED harmonics is incoherent and can achieve extremely high frequency reaching gamma-ray band. Based on such mechanism, we present the generation of a gamma-ray train by impinging a linearly polarized laser on a solid target. The simulation shows that the period of the train is $0.5T$ and the peak intensity of gamma-ray pulse is $1.4 \times 10^{22}$ W/cm$^2$ in our case. By tracing the typical electron trajectories, the simulation result is identified by numerical calculation, which rightly reproduces the gamma-ray train and the energy spectrum. As expected, the oscillating electrons in the layer are the sources of radiation and the corresponding energy spectrum supports the conclusion that the radiation is incoherent and is more likely to act as a particle beam. Further study shows that under such an extremely strong electromagnetic field, ions will absorb energy from laser pulse and gain a velocity. This reduces the intensity of the gamma-ray radiation and the movement of ions will enhance the radiation at forward direction and suppress the radiation from backwards. In addition, multi-dimensional effects are also investigated and the 2D simulation shows that the generation of gamma-ray train still exists. The mechanism of 'QED harmonics' reported in this work is of crucial importance for the physics of laser-matter interactions in the strong filed domain in showing that the energy conversion from laser to extremely high frequency radiation is by quantum and in a random way. A potential way to produce high-quality gamma-ray pulses is provided and will be useful for ultra-fast detection and other applications. The realization of high-power lasers of $10^{23}$ W/cm$^2$ in the foreseeable future will open opportunities to explore the charming features in the laser-matter interactions and this result can be verified in experiments.

This work was supported by the Ministry of Science and Technology (Grants No. 2011CB808104), and the National Natural Science Foundation of China (Grants No. 61221064, No. 11374319, No. 11125526, No. 11335013, No. 11127901).

*To whom all correspondence should be addressed: bfshen@mail.shcnc.ac.cn,

zhxm@siom.ac.cn